\documentclass[preprint,proceedings]{rmaa}

\newcommand{\D}{\discretionary{}{}{}}

\SetYear{2009}
\SetConfTitle{The Interferometric View on Hot Stars}

\title{Prospects for the study of dust making Wolf-Rayet binaries with the VLTI-Spectro-Imager (VSI)}

\author{
  M. De Becker\altaffilmark{1}, M. Filho\altaffilmark{2} and T. Harries\altaffilmark{3}}

\altaffiltext{1}{Institut d'Astrophysique et G\'eophysique, University of Li\`ege, 17 all\'ee du 6 Ao\^ut, B\^at. B5c, 4000 Li\`ege (Sart-Tilman), Belgium
  (debecker{}@astro.\D{}ulg.\D{}ac.\D{}be)}
\altaffiltext{2}{CAUP, University of Porto, Portugal}
\altaffiltext{3}{School of Physics, University of Exeter, United Kingdom}

\suppressfulladdresses

\listofauthors{M. De Becker, M. Filho, T. Harries}

\indexauthor{De Becker, M.}

\addkeyword{Stars: Binaries}
\addkeyword{Stars: Early-type}

\begin{document}

\maketitle 

\section{The VLTI-Spectro-Imager}
The VLTI-Spectro-Imager (VSI) is a 2nd generation VLTI instrument under development (Malbet et al.\,2008). The main improvement of VSI with respect to previous interferometric instruments (e.g. AMBER) is a significant enhancement of the imaging capability, thanks to the combination of 6 beams resulting in a significant spatial frequency gain. 

The science input will be further improved using the spectral dispersion of VSI, likely to reveal wavelength dependent structures in targets (e.g.. spectral lines produced in stellar winds of hot stars), and by high dynamic range observations that are expected to improve our understanding of many classes of objects (e.g. shells around evolved massive stars). In the context of the activities of the science group of the VSI consortium, many science topics have been considered, including a few devoted to massive stars (see Garcia et al.\,2008).

\section{Dust making Wolf-Rayet stars}
Among the most attractive targets for VSI are dusty shells around Wolf-Rayet (WR) binaries. Most of the carbon rich WR systems (WC-type + O) are known to display an infrared excess emission related to the presence of dust. The dust shells absorb stellar radiation and re-emit the energy in the infrared. A few of them have been resolved in pinwheel nebulae, defined by rotating spiral dust shells. Dust formation is strongly related to binarity. The higher density associated with wind-wind collision regions favors indeed dust condensation, with enhanced formation close to periastron, and the dust flows at the opposite of the Wolf-Rayet star whose wind is stronger than that of the O-star in the binary (Williams et al.\,1985,1994). 

Examples are provided in Fig.\,1 for the reconstruction of K-band synthetic images for an inclination of 0$^\circ$. The left panel gives the synthetic image, and the right panel provides the reconstructed image assuming that 6 telescopes are used simultaneously. We see that the central part of the nebula is recovered from the image reconstruction. The faintest parts at the outskirt of the nebula are not recovered. However, high dynamic range observations performed by cumulating observations of the same target are expected to improve substantially the quality of images obtained with VSI. For details on the image reconstruction, we refer to Filho et al.\,(2008).

\begin{figure}[!t]
  \includegraphics[width=\columnwidth]{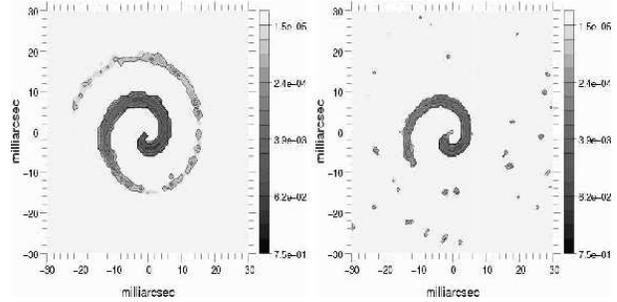}
  \caption{Left: synthetic image of a pinwheel nebula surrounding a WR binary. Right: reconstructed VSI image.}
  \label{pw}
\end{figure}

The detailed investigation of the near infrared radiation from dust making WR binaries will allow us to probe the properties of dust particles (typically their size), along with their production rate. The high angular resolution of the VLTI is expected to help us to study the distribution of the properties of dust across the pinwheel nebula, and to potentially discover new nebulae of the same kind through a first systematic imaging of WC-type binaries. The monitoring of such systems is expected to allow to investigate the dust formation mechanism as a function of the orbital phase. Changes are indeed expected if one is dealing with eccentric binaries. In addition, it should be noted that the shape of the nebula is intimately related to the parameters of the orbit.

\end{document}